# WIRELESS SENSOR NETWORKS LOCALIZATION ALGORITHMS: A COMPREHENSIVE SURVEY.

Asma Mesmoudi[1] , Mohammed Feham[1], Nabila Labraoui[1]

[1] STIC Laboratory, Departemet of telecommunication, University of Tlemcen, Algeria

*ABSTRACT*

*Wireless sensor networks (WSNs) have recently gained a lot of attention by scientific community. Small and inexpensive devices with low energy consumption and limited computing resources are increasingly being adopted in different application scenarios including environmental monitoring, target tracking and biomedical health monitoring. In many such applications, node localization is inherently one of the system parameters. Localization process is necessary to report the origin of events, routing and to answer questions on the network coverage ,assist group querying of sensors. In general, localization schemes are classified into two broad categories: range-based and range-free. However, it is difficult to classify hybrid solutions as range-based or range-free. In this paper we make this classification easy, where range-based schemes and range-free schemes are divided into two types: fully schemes and hybrid schemes. Moreover, we compare the most relevant localization algorithms and discuss the future research directions for wireless sensor networks localization schemes.*

*KEYWORDS*

*localization, WSN, anchor node, range-based methods, range-free methods, hybrid-based methods.*

## 1. INTRODUCTION

Recent developments in MEMS IC technology and wireless communication have made possible the use of large networks of wireless sensors for a variety of applications including process monitoring, process control [1]. A Wireless sensor network (WSN) is formed by hundreds of small, cheap devices called sensors which are constrained in terms of memory, energy and processing capacities [2]. These sensors are deployed to sense the physical characteristics of the world, such as temperature, light and pollution. WSNs are expected to be solution to a wide range of applications such as monitoring, natural disaster relief, patient tracking, military target and automated warehouses. In many of these applications, location awareness is useful or even necessary [3]. Indeed, without knowing the position of sensor node, collected data is valueless. The localization of sensors can be implemented by different manners. A simple solution is to equip each sensor node with a GPS receiver that can precisely provide the sensor nodes with their accurate position [3]. However, adding the GPS to all nodes in the wireless sensor network is not practical because of high cost, high power consumption and environment constraint [4]. In addition, the GPS fails in indoors applications, under the ground, or dense forest.

Self-localization is an alternate solution of GPS, in which sensor nodes can estimate their position by using various localization discovery protocols. These protocols share a common characteristic : most of them use a few special nodes, called beacon nodes, which are assumed to know their own locations (through manual configuration or GPS receivers) [5]. These beacon nodes (also referred to as anchor nodes, seeds, references or landmarks) provide position information, in the form of beacon messages, for the benefit of non-beacon nodes, well known as blind nodes (also referred to as unknown nodes, dumb nodes or target). Blind nodes can utilize the position information of multiple nearby beacon nodes to estimate their own positions [6].

Almost all existing localization schemes consist of two phases: 1) distance/angle estimation; 2) position computation. In distance/angle estimation, the most common range measurement techniques used to estimate distance or angle between two sensor nodes are TDOA (Time Difference Of Arrival , TOA (time of arrival) , RSSI (Received Signal Strength Indicator), AOA (angle of arrival) and Hop-count. In position computation, the position of the unknown node is estimated based on the available information of distance or angle and positions of references nodes. The commonly used techniques include lateration, triangulation, bonding box, probabilistic approach and fingerprinting.

Several surveys are proposed in the literature, in which, the authors describe different taxonomies that classify the localization schemes for sensor networks based on several distinct criteria such as: dependency of the range measurements (i.e. range-based localization or range-free localization); distributed or centralized position computation; with or without an infrastructure (anchor based localization or anchor-free localization).

According to the dependency of range measurement techniques, localization algorithms can be classified into two main categories: range- based schemes and range- free schemes. Almost of surveys, as the case of surveys [7][8] ,they completely ignore the recently emerging solutions based on hybrid systems, which combine different methods based on connectivity information and/or range measurement techniques. Moreover, there are  either range-based or range-free schemes. Clearly, it is infeasible ,if not difficult, to classify the hybrid schemes as range-based or range- free.

In this paper, we make the classification of localization algorithms based on the dependency of the range measurements easy with a focus on the hybrid algorithms, this classification is proposed also to help in comparing localization schemes. For example, hybrid- range- based schemes perform more accurate than the fully range- based schemes. And a comparative study of existing localization algorithms from different perspectives is discussed in detail.

The rest of this paper is organized as follow. In section 2, classification of localization algorithms is given. In section 3, comparative study of localization algorithms is discussed. Section 4 concludes the paper and outlines future possible research.

## 2. CLASSIFICATION OF LOCALIZATION ALGORITHMS

According to the dependency of range measurements, the existing localization schemes can be classified into two major categories: the range-based approaches and the range-free approaches.
The range-based schemes are based on using range measurement techniques for location estimation. The range-free schemes ignore the using of range measurement techniques. Thus, in order to estimate the location of unknown nodes, these schemes are based on the use of the topology information and connectivity, i.e.," who is within the communication range of whom [9]. Furthermore, there are schemes that combine different methods based on connectivity information and/or range measurement techniques. Clearly, it is difficult, if not infeasible, to classify these schemes as range-based or range- free. In this paper, range-based schemes and range-free schemes can be divided into two sub-categories: fully schemes and hybrid schemes, as shown in figure 1. This classification is based on the dependency of the methods used and have a direct impact on the estimation of unknown node location . For example, there is schemes use at the same time range- based and range- free mechanisms (therefore be either considered as range-based or range- free schemes). According to the definition of range- based schemes mentioned above, if the schemes use range measurement techniques they are considered as range- based schemes ,and if not they are considered as range- free schemes. Thus the hybrid schemes that combine range- based and range- free mechanisms are considered as range- based schemes.

Range- based or range- free schemes may or may not use anchor nodes, i.e., anchor based or anchor free.

The anchor-free schemes do not assume any node positions are initially known. While, the anchor-based schemes need some nodes aware of their positions called anchor nodes to provide geographic information to unknown nodes to localize . A promising method is to use mobile anchor node instead of static anchor nodes [10]. A mobile anchor node is aware of its position, and moves in sensor area and broadcasts its current position periodically to generate a number of virtual anchor nodes. The unknown sensor nodes estimate their locations by measuring the geographic information (e.g., distance or angle) of the virtual anchor nodes.

In next section of this paper, we analyze and summarize the typical localization schemes of each category

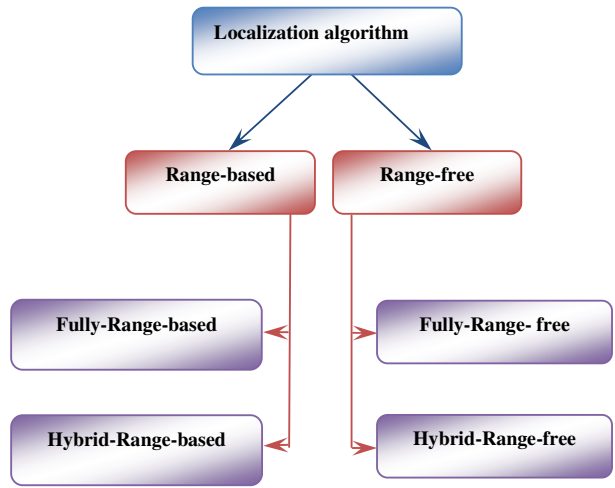

Fig.1. Classification of localization schemes for sensor networks

## 2.1. Range-free localization algorithms

In order to estimate the location of unknown node , This category is based on the use of the topology information and connectivity, i.e.," who is within the communication range of whom", [9]. According to the manner that location of unknown node is obtained. The range-free schemes can be further divided into two types: fully-range-free and hybrid-range-free schemes.

### 2.1.1. Fully-range-free localization algorithms

This type of algorithms uses only one method based on connectivity and/or information of topology. Among the main approaches in the literature, we quote them below:

**A. Anchor Based Approaches**

The commonly fully-range-free schemes based on anchor nodes include centroid, DV-hop, APIT, and they are described below:

Bulusu et al. proposed in [11] Centroid localization Algorithm (CA). It is the most basic scheme that uses anchor beacons, containing location information $(x_i, y_i)$ to estimate node position. After getting these beacons, the unknown node use the following centroid formula to computes its location [11]:

$$(\hat{x}, \hat{y}) = \left( \frac{\sum_{i=1}^{n} x_i}{n}, \frac{\sum_{i=1}^{n} y_i}{n} \right) \quad (1)$$

Where n is the number of the anchor nodes $A_i$, and $(x_i, y_i)$ are their coordinates, and $(\hat{x}, \hat{y})$ is the estimated coordinate of blind node. After that, the unknown node becomes an anchor node and broadcasts packets to the vicinity.

This method of localization is very simple, economic and eases of implementation. However, localization accuracy is vulnerable to be affected by density of anchor nodes and deployment of nodes. Besides, this algorithm focuses only on node for 2D networks, thus the authors in [12] propose a Novel Centroid localization Algorithm (NCA) for three dimensional WSNs based on the centroid theorem of coordinate-tetrahedron [13][14].

Niculescu and Nath proposed DV-Hop algorithm [15, 16], that is based on hop distance vector algorithm and the estimation of unknown node location comprise of three steps:

First, it apply a classical distance vector exchange. So, through this mechanism, all nodes in the network get minimal hop-count to every anchor nodes.

Second, once an anchor node receives the hop count value to other anchor nodes, it computes an average size for one hop by the follow formula, which is then deployed to the nodes in its neighborhood.

$$\text{HopSize}_i = \frac{\sum_{j \neq i} \sqrt{(x_i - x_j)^2 + (y_i - y_j)^2}}{\sum_{j \neq i} h_{ij}} \quad (2)$$

Where $(x_i, y_i)$, $(x_j, y_j)$ are coordinates of anchor i and anchor j, $h_{ij}$ is the hops between them. When receiving the average hope size, the unknown node calculates the distance to the anchor nodes as follow:

$$d_i = \text{HopSize}_i \times L_i \quad (3)$$

Where, $L_i$ is the hop value between i[th] anchor node and unknown node.

Third, when the unknown nodes get three or more distance information from beacon nodes, they can use trilateral method to estimate their coordinates.

The DV-hop localization scheme is characterized by its simplicity, ease of implementation, and the fact that it does not depend on range measurement error. But on the other hand this scheme will only employ for isotropic networks, that is, when the properties of the graph are the same in all directions [16]. Besides, this scheme has bad localization accuracy. Therefore some improved DV-HOP methods have been reported to enhance the localization accuracy. In [17], a new computational model is developed by considering the relationships between the hop-distances and the communication ranges, the authors denote this proposal scheme as CDV-Hop.

The authors in [18] propose a Selective ANchors node Localization Algorithm (SANLA), where the unknown node chooses three anchors which have the best precision for localization from all the anchor nodes it received. In [19], each anchor node is given a weight which is production of hop size weight and position weight. In [20], authors add correction when computing the distance between the anchor nodes and unknown nodes.

He et al. proposed APIT [21], an Approximate Point-In-Triangulation test based on the principle of divide the whole network into triangular regions. theses triangular regions made up of vertices formed by all the possible set of connecting three neighboring anchor nodes . The unknown node apply a test to determine whether it is inside\outside the triangle formed, the test is repeated for all connecting three anchor nodes heard by unknown node, and the location is estimated as the center of gravity of the triangles' overlapping region.This algorithm is little influenced by environmental factors and requires a low hardware, which is a widely accepted positional mechanism for WSNs. however, the localization accuracy in the APIT method is affected by a node's presence whether it is within the triangular regions or not. In [22], an improvement scheme decreases the probabilities of In-To-Out Error and Out-To-In Error from the traditional APIT scheme.Besides, the practical environment is always a three dimensional situation and APIT has a bad accuracy in this case. [23] proposed an improved APIT-3D scheme based on volume-test, named Volume Test Approximate Point-In-Triangulation testthree-dimension (VT-APIT-3D) algorithm.

However, the main drawback of APIT algorithm is requiring more anchor nodes than the average number of anchor used in localization.We noted that the APIT algorithm does not make any assumption about the correlation between absolute distance and radio signal strength; hence, we consider it as a fully-range- free algorithm.

The scheme described in [24], is another area based range-free localization. The authors assume that the anchor nodes are equipped with directional sectored antennas, while the remainder nodes are equipped with omnidirectional antennas. In this scheme, a sensor determines a node location as a center of gravity of the overlapping region based on the information transmitted by the anchor nodes. This scheme is unique in its secure design. It can deal with diffrent types of attacks including Sybil and wormhole attacks [25].

The scheme described in [26] is also another area based range-free Localization Algorithm which uses a mobile beacon instead of static beacon nodes called the Azimuthally Defined Area Localization (ADAL) method. This scheme uses a mobile beacon with a rotary directional antenna to send message in a determined azimuth periodically, and an unknown node uses the centroid of the intersection area of several beacon messages as its position [10]. However, the shortest trajectory that traverses each mobile beacon is hard to find.

**B. Anchor Free Approaches**

In [27] the authors propose a centralized range-free algorithm called MDS-MAP. Which is based on multidimensional scaling (MDS), it estimates position of the unknown node depending on the basic information of the nodes in the communication range.

MDS-MAP scheme is capable to form relative maps that represent the relative positions of sensor nodes when there are no anchor nodes [28]. When the positions of a adequate number of anchor nodes are known (3 anchor nodes for 2-D localization and 4 anchors for 3-D), the absolute coordinates of all nodes in the map is estimated.

MDS-MAP generates the most accurate location information among range-free techniques. However, it suffers from the following drawbacks: The time complexity is high, large bandwidth and computation are required to estimate locations when there are a large number of sensors. Thus, in [29] an improved algorithm has been proposed (IMDS-MAP), where a distributed positioning algorithm has been implemented by clustering.

### 2.1.2. Hybrid-range-free localization algorithm

This type of algorithms combines between different methods based on topology and/or connectivity for location estimation.

**A. Anchor based approaches**

Xinhua and Zhongming proposed in [30] an Iterated Hybrid Localization Algorithm (IHLA) based on the centroid scheme and DV-Hop scheme. When each unknown node computes its initial coordinates by using the centroid scheme, it estimate again the distances among each unknown node to the beacon nodes based on the DV-Hop scheme. After that, Taylor Series Expansion (TSE) algorithm is used to estimate coordinates of each unknown node.

The proposed algorithm has better localization accuracy compared with centroid scheme and DV-hop scheme. However, this hybrid algorithm is more complex and needs more computing time.

**Summary**

Range- free approaches are expected to be alternative solutions to range-based approaches. They do not require any extra hardware, because they do not rely in any distance measurements. The main advantages of range-free approaches are its simplicity and low cost. However, the localization error is highly dependent on the density of nodes, on the number of beacon nodes and on the network topology. They are suitable for applications where location accuracy is less critical.

Hybrid approaches, which combine different range-free methods taken the advantages of each one, these schemes can achieve a performance improvement over that based on a single method. However, they are more complex and need more computing time.

### 2.2. Range-based localization algorithm

This category of algorithms is based on using range measurement techniques for location estimation. According to the manner of using the range measurement techniques, this category can be divided into two types: fully-range-based and hybrid-range-based localization algorithm. Both of them are even anchor based or anchor free.

### 2.2.1. Fully-range-based localization algorithm

These algorithms use only one type of range measurement techniques to estimate the distance or angle between nodes. The unknown node can estimate its coordinates by using one of the methods discussed in section 2.2.

The fully-range- based localization may or may not require anchor nodes as describe below.

**A. Anchor based approaches**

In the anchor based approaches, there are different fully-range-based localization algorithms based essentially on the Received Signal Strength Indicator, Angle of Arrival and Time to Arrival.

The proposed schemes in [31, 32, 33, and 34] are based on the estimation distances between neighboring sensor nodes from the received signal strength measurement.

The Received Signal Strength Indicator (RSSI) is based on the physical fact of wireless communication that theoretically, the signal strength is inversely proportional to the squared distance between a pair of sensor nodes. A known radio propagation model is used to convert the received signal strength into distance. In RSSI techniques, either empirical or theoretical models are used to translate signal strength into distance [35].

Among the range-based measurement techniques, the RSSI technique is the most common techniques, cheapest and simplest, since its low cost because it does not require additional hardware (e.g. infra red or ultrasonic). However, extending a RSSI-based technique for 3D localization can introduce higher complexity in computational cost and location accuracy. Thus, the authors in [32] propose a COmplexity-reduced 3D trilateration Localization Approach (COLA) based on RSSI values. this proposed scheme simplify the positioning process by decreasing 3D computation to 2D computation.

Furthermore, in real environment the RSSI is very susceptible to noise and obstacles, particular for indoor environment [36]. It should consider errors in the measured values, which can be obtained from multi-path propagation, fading effects and reflection [37].On the other hand, An RSSI-based scheme therefore requires more data compared with other methods to achieve higher accuracy [38]. However, when a large amount of data was collected an increase in traffic and in the energy consumption of sensors was occurred and this will decrease the lifetime of sensor networks. Thus the authors in [33] proposed an Indoor LOcalization system using RSSI measurement of wireless sensor network (ILOR) based on zigbee standard to decrease the amount of data collected by the sink and prolong the lifetime of the sensor networks.

The shcemes considered in [39] [40] [41] [42], [43] [44], are based on Angle Of Arrival measurements (AOA measurements), which is known also as Direction Of Arrival (DOA)
In [40], two algorithms were proposed DV-Radial and DV-Bearing, the AOA ability provides for each node bearings to neighboring nodes with respect to a node's own axis. A radial is the angle under which an object is seen from another point or simply a radial is a reverse bearing. AOA based schemes are described where sensor nodes are transmitting their bearings with respect to beacon nodes ( i.e. nodes which are know their own coordinates and orientations). In order to estimate the orientation, a device known as The Cricket Compass uses ultrasound with respect to a number of ceiling mounted beacons [45].

Unfortunately, the methods considered in [40] and [43] need a strong cooperation between neighbor sensors, and they are prone to error accumulations.

Other approaches as considered in [42], proposed an Angle-of-arrival Localization based on antenna ARrays for wireless sensor networks (ALAR). The AOA measurements are driven from the measurements of the phase differences in the arrival of a wave front. It usually needs a large receiver antenna (relative to the wavelength of transmitter signal) or antenna array. This approach works quite well for high average Signal-to-Noise Ratio (SNR) but in the existence of strong multipath signals and/or co-channel interference this approach may fail [39].
The advantage of this scheme is the high accuracy. However, it is limited by directivity of the antenna, by shadowing and by multipath reflections, and the main disadvantage is the additional hardware requirement.

The algorithms considered in [46] [47] [48] [49][ 50], are based on time of arrival (TOA) which known also as time of flight (TOF) [46], where the distance between transmitter and receiver obtained through multiplying the propagation time of the signal by its propagation speed [48].
This technologies is also used by the global positioning system (GPS) [51], among the limits of TOA is the necessary to have a synchronized transmitter and receiver, the synchronization adds cost and complexity to the WSN [52].Thus, Chen et al. proposed in [47] a Mobility-Assisted Node localization Based on TOA measurements (MABT) without time synchronization in WSN.

Furthermore, TOA method is most suitable for under water and underground exhibiting low propagation speeds [53, 54]. Lee et al. proposed in [46] TOA based sensor Localization in Underwater Wireless sensor networks (TLUW). This approach is applicable to a number of applications, including underwater target tracking, seismic monitoring, equipment monitoring, leak detection, etc.

The algorithms considered in [4] [55-57], are based on Time Difference Of Arrival (TDOA) measurements, There are mainly two ways to obtain TDOA in wireless sensor networks [58]:
In the first way, TOA can be measured from an unknown node to two different anchor nodes, and calculates the time difference. This generates a hyperbolic curve aim attention at the two beacon nodes on which the unknown node must lie. When three or more than three hyperbolic curves are getting with three or more than three propagation time differences, the unknown node location is the unique intersection point of these hyperbolic curves. In the same way, Xiao et al. proposed in [55] a Research of TDOA based Self-localization Approach in Wireless Sensor Network (RTSA).which achieves average value of time difference by rolling average to decrease the error of measurement . However, time synchronization of anchor nodes and time synchronization of unknown nodes still are required.

In the second way, two transmission mediums of very different propagation speed are accessible, that's mean two different signals, for example ultrasound/ acoustic and radio signals. In the same way, Savvides et al present in [4] a novel location discovery approach, which they call AHLoS (Ad-Hoc Localization System), for wireless sensor network.

To estimate distance between sender and receiver and for simple case when the transmission of the two difference signals is simultaneous, and time delay equal to zero, thus the receiver use simple formula to compute the distance between its self and the sender, as follow [4]:

$$d = s_{sound} \times (t_{sound} - t_{radio}) \quad (4)$$

Where, $s_{sound}$ is the speed of sound waves and $t_{sound}, t_{radio}$ are travel times of acoustic and radio signals respectively.

The main drawback of this technique is that requires additional hardware in every node in order to transmit and receive the second signal (acoustic/ultrasound signal), that makes system costly (expensive and energy consuming).besides, the ultrasound signal can be stopped by obstacles.

**B. Anchor free approaches**

Among the common fully-range-based localization algorithms that don't require anchor nodes, we quote them below:
 ABC (Assumption Based Coordinates) algorithm [59] is based on RSSI measurements to determine the inter node distances. In order to satisfy the inter-node distances ,this scheme first chooses four in range sensor nodes and assigns them coordinates. The coordinates of other nodes are incrementally calculated using the distances from at least four nodes with already calculated coordinates.

The ABC algorithm is relatively simple and does not require complicate calculation, but the localization accuracy is poor, especially for widespread networks. As with all incremental algorithms, error propagation is cumulative which results in poor coordinate assignment. In particular, positioning accuracy decreases moving away from the node which started algorithm and in real network the complete graph realization is not always guaranteed. Furthermore, if measurements are corrupted by noise, the algorithm can lead to incorrect nodes localizations.

In [60] the authors proposed a RObust Distributed network Localization with noisy range measurements (RODL) for locating nodes in a sensor network in which the node measures distances to neighboring using the time difference of arrival (TDOA). In particular, the authors consider how the measurement noise can cause incorrect realization of node displacement. In this approach, each node becomes the center of a cluster and computes the relative location of its neighbors which can be absolutely localized. Once is done, an optional optimization can be deployed to refine the localization of clusters. Cluster stitching technique is used to obtain a coordinate assignment for all the nodes, within a general coordinates system.

The advantage of this approach is that successfully localized nodes in sensor network with noisy measurements, using no anchor nodes. And as its limitation is that this approach may be unable to localize a useful number of unknown nodes under condition of high measurement noise or low node connectivity. To get over these difficulties, [61] proposes an enhancement of the algorithm presented in [60] that uses a new quadrilateral robustness test.

**A.2.2 Hybrid-range-based localization algorithm**

 Hybrid-range-based schemes combine different distance or angle estimation methods. This combination may be between the range measurement techniques only or between range measurement techniques and connectivity methods. Among the main approaches that exist in the literature, we quote them below.

*A. Anchor based approaches*

Among the hybrid localization algorithms that require anchor nodes, are described below:

The authors in [62] propose a distributed AOA aided TOA Positioning Algorithm in mobile wireless sensor networks (ATPA). This scheme performs location estimation in three steps. In the first step, since the movement of the unknown node presents differences between arrival times of beacon nodes, in order to modify the TOA measurements the aided AOA information may be applied, which can be used to estimate the unknown location. In the second step, a geometrical positioning with particle filtering is applied to compute the location of unknown node from state equations [63]. In the third step, to solve the localization adjustment problem an adaptive fuzzy control is used [64]. In ATPA scheme  an adaptive flexibility and robust enhancement are provided  in the computation with moderate noisy measurements.

Bishop et al. proposed in [65] Exploiting geometry for improved hybrid AOA/TDOA-based localization (EATL). This scheme is based on the combination techniques of bearing (angle of arrival AOA) and time difference of arrival TDOA techniques. First, each beacon node measures the target bearing (positive counter-clockwise from the x-axis) which is equal to the sum of the true bearing and an error. Moreover, each anchor node measures the time of signal arrival, which is equal to the sum of true time of signal arrival and an error. Then the time-difference between the arrival times at anchor nodes gives the distance–difference measurements. After that, the authors formulate a constraint function that is used in a constrained optimization process; this process estimates the maximum likelihood measurement errors such that the final solution satisfies the proposed constraint which captures the underlying geometry. Thus the maximum likelihood measurement errors when subtracted from the measurements give values of bearing and distance-difference that permit a consistent location estimate [65]. The proposed algorithm is inherently more robust to initialization procedures than the traditional maximum likelihood estimation techniques.

Desai et al. proposed in [66] Fusion of RSSI and TDOA measurements from wireless sensor network for robust and accurate indoor Localization (FRTL) that combines RSSI and TDOA measurements in location system. This algorithm uses TDOA as a primary distance estimation scheme for localization. It collects and trains RSSI data with the associated known distances, in parallel. Trained distances based on appropriate RSSI value can then replace any missing TDOA measurements. In presence of acoustic noise when TDOA communication is unavailable, the algorithm can also use trained distances in place of all missing TDOA measurements [66].

The performance of the proposed scheme FRTL is much better in term of distance determination and localization compared to techniques, which relies solely on TDOA or RSSI measurements.

In [67] a Hybrid TOA/RSSI Localization algorithm (HTRL) is proposed, which utilized TOA and RSSI measurements in the location system. The algorithm, using only three TOA range measurements, does not require the knowledge of LOS (line of site) /NLOS (non line of site) conditions. The scheme encourages the objective function from the geometrical relationships of the beacon nodes and TOA range circles. It uses the RSSI and a pre-established path loss model, which is assumed to be well approximating the propagation conditions, to discriminate between LOS or NLOS range measurements. Based upon the result of the hypothesis testing, the weight factors are assigned [67].

The weight factors are used in describing the credibility of the TOA range measurements, and to determine the values of weight factors, the relationship between RSSI and distance is utilized.
The proposed hybrid TOA/RSSI location algorithm performs better than other algorithms and increases performance of localization under severe NLOS.

Blumenthal et al. proposed in [68] Weighted Centroid localization Algorithm (WCA). This scheme takes into account RSSI metrics and used it in centroid scheme. The anchor nodes first broadcast packets (anchor's id, location, transmission power) to its neighborhoods. Then, the unknown node receives the packets transmitted by the different anchor nodes and selects the anchor nodes (at least three anchor nodes) with larger RSSI (anchors closer than others). Next, the unknown node computes the weight of each anchor node using the obtained RSSI measurements and uses it to determine its position.

This proposed scheme utilizes weights to attract the estimated location to close anchor nodes provided that coarse distances are available, where it enhances the performance in term of localization of centroid method without any additional hardware cost and negligible energy consumption. The authors in [10] proposed a Three-Mobile-beacon assisted Weighted Centroid localization scheme (TMWC), where these beacons preserve a special formation while traversing the network deployment area, and broadcast their positions periodically. The unknown nodes estimate the distances to these beacons by using RSSI measurements, and utilize centroid localization scheme to calculate its position.

Tian et al. proposed in [69] a RSSI-based DV-hop localization algorithm (RDV-Hop), which incorporates RSSI and DV-hop to implement localization together. First, each anchor node estimates an average size for one hop. Then, the anchor nodes broadcast RSSI packets in the network, and the unknown node which is one hop away from anchors use RSSI method to calculate the distance between itself and the neighbor anchors. On the other side, the same unknown node uses the average size for one hop which has known before to calculate the distance between itself and the other anchor nodes (anchor nodes have not one hop distance away from unknown node). Once the unknown node get distances from anchor nodes, it can locate itself by triangulation method.

This algorithm improves the localization accuracy compared with the DV-hop algorithm, but the limitation of this algorithm is that only neighbors of anchor can obtain RSSI distances.

*B. Anchor free approaches*

Among the hybrid localization algorithm that don't require anchor nodes are presented below:

Magnani and Leung. proposed in [70] Self-Organized, Scalable GPS-free Localization of Wireless Sensors (SSFL), that fuses the Time Difference of Arrival (TDOA) and the Angle of Arrival (AOA) measurements for estimating the ranges and angles among sensor nodes. In the beginning, each sensor uses the estimates (ranges and angles) obtained locally to build a network coordinate system. Then, the local coordinate system is corrected with respect to the reference coordinate system. After that, the global coordinate system for positioning is defined.

Priyantha et al. proposed a concurrent algorithm in [71] called Anchor Free Localization algorithm (AFL), where all the nodes calculate and refine their coordinate information in parallel. The AFL algorithm begins with an initial coordinate assignment based on the connectivity between nodes and concurrently is followed by using a mass-spring optimization to correct localization errors which is based on more accurate inter-node distance, measured using TDOA. Instead of incremental algorithms, AFL performs much better, even for networks with small connectivity. Furthermore, AFL error propagation is small.

The main drawback is that the initial position estimation will make a greater impact on the AFL algorithm and according the simulation result in [71] a pure mass-spring algorithm (i.e. algorithm without fold-freedom) does not work without good initial position estimates. Furthermore, if measurements are corrupted by noise the algorithm can lead to incorrect nodes displacements.

*Summary*

The localization schemes based on angle of arrival (AOA) and propagation time (TOA and TDOA) measurements can achieve better accuracy compared to the schemes based on RSSI measurements, because the environmental factors affect the amplitude of the radio signal. However, this accuracy is achieved at the expense of higher cost equipment.

Most range-based schemes are not appropriate to low density networks. The connectivity failures due to higher distance between nodes, are leading to hinder the computation of distance measurements .

However, hybrid localization schemes can achieve a performance improvement over that based solely on a measurement kind, because measurement noise for different kinds of measurements arrives from different sources. Therefore errors in the position estimate for each measurement kind are at least partially independent . This independence between different measurements kinds can be exploited by data fusion techniques to create estimators that have better accuracy than estimators based on single measurement kinds. However, in hybrid localization scheme the time complexity is high, and large computation are required to estimate locations compared with fully localization scheme.

## 3. PERFORMANCE COMPARISON

A comparative study is presented in this section for range- based and range-free algorithms based on the follow performance parameters: network assumptions (deployment, Node density ,existence of obstacle, existence of Anchor node, nodes mobility and mobile assisted), localization process (range estimation, range combination, computational model and localization coordinates), and design goal (scalability, overhead and accuracy). We summarize this comparison in table 1. In the following we will discuss the metrics used to evaluate design goal: accuracy, scalability and overhead.

| | Net assumptions | | | | | | Localization process | | | | Design goal | | | | |
|---|---|---|---|---|---|---|---|---|---|---|---|---|---|---|---|
| | | | | | | | | | | | | Overhead | | | |
| **Fully-range-free** | De | Nd | Ob | An | Nm | Ma | Rae | Rac | Com | Lc | Sca | Cm | Cp | Hc | Ac |
| **CA [11]** | Both | L | Y | Y | N | N | Conn | Centro | Dist | 2D | Y | L | L | L | L |
| **NCA [12]** | R | L | Y | Y | N | N | Conn | Centro | Dist | 3D | Y | L | M | L | L |
| **CDV-Hop [17]** | R | H | - | Y | N | N | Conn | Multi | Dist | 2D | N | H | L | L | M |
| **ADAL [26]** | R | H | - | N | N | Y | Conn | Centro | Dist | 2D | Y | L | L | L | M |
| **APIT [21]** | Both | H | Y | Y | N | N | Conn | Centro | Dist | 2D | Y | L | L | L | M |
| **VT-APIT [23]** | R | M | - | Y | N | N | Conn | Centro | Dist | 3D | Y | L | L | L | M |
| **MDS-MAP [27]** | R | L | - | N | N | N | Conn | Multi | Cent | 2D | N | H | H | L | H |
| **IMDS-MAP [29]** | R | L | - | N | N | N | Conn | Multi | Dist | 2D | Y | M | M | L | H |
| **Hybrid-range-free** | De | Nd | Ob | An | Nm | Ma | Rae | Rac | Com | Lc | Sca | Cm | Cp | Hc | Ac |
| **IHLA [30]** | R | M | - | Y | N | N | Conn | Centro TSE | Dist | 2D | N | H | M | L | H |
| **Fully-range-based** | De | Nd | Ob | An | Nm | Ma | Rae | Rac | Com | Lc | Sca | Cm | Cp | Hc | Ac |
| **ILOR [33]** | R | M | - | Y | N | N | RSSI | MLE | Cent | 2D | N | H | H | L | H |
| **COLA [32]** | R | M | Y | Y | N | N | RSSI | Trilat | Dist | 3D | Y | L | M | L | M |
| **DV-bearing [40]** | R | H | Y | Y | Y | N | AOA | Triang | Dist | 2D | Y | H | L | L | H |
| **ALAR [42]** | R | M | Y | Y | N | N | AOA | LSE | Dist | 2D | Y | L | L | H | H |
| **MABT [47]** | R | M | Y | N | N | Y | TOA | LSE | Dist | 2D 3D | Y | L | L | L | H |
| **RTSA [55]** | U | - | Y | Y | N | N | TDOA | LSE | Dist | 2D | Y | L | L | L | H |
| **AHLoS [4]** | U | M | N | Y | N | N | TDOA | Multi | Dist | 2D | Y | L | L | H | H |
| **ABC [59]** | R | M | Y | N | N | N | RSSI | Triang | Dist | 2D 3D | N | L | L | L | L |
| **RODL [60]** | Both | M | Y | N | Y | N | TDOA | Trilat | Dist | 2D | Y | M | M | H | M |
| **Hybrid-range-based** | De | Nd | Ob | An | Nm | Ma | Rae | Rac | Com | Lc | Sca | Cm | Cp | Hc | Ac |
| **ATPA [62]** | R | H | Y | Y | Y | N | AOA TOA | BPF | Dist | 2D | Y | M | M | H | V.H |
| **EATL [65]** | Both | L | Y | Y | N | N | AOA TDOA | MLE | Dist | 2D | N | M | H | H | V.H |
| **FRTL [66]** | - | - | Y | Y | N | N | TDOA RSSI | LSE | Cent | 3D | N | H | H | H | H |
| **HTRL [67]** | U | L | Y | Y | Y | N | TOA RSSI | MLE | Dist | 2D | Y | M | M | L | H |

| | | | | | | | | | | | | | |
|---|---|---|---|---|---|---|---|---|---|---|---|---|---|
| **SSFL [70]** | Both | - | - | N | N | N | TDOA AOA | Triang | Dist | 3D | Y | M | H | H | V.H |
| **WCA [68]** | U | - | Y | Y | N | N | RSSI Conn | Centro | Dist | 2D | Y | L | M | L | H |
| **RDV-hop [69]** | R | M | - | Y | N | N | RSSI Conn | Trilat | Dist | 2D | N | H | L | L | H |
| **AFL [71]** | R | L | - | N | Y | N | TDOA Conn | Trigon | Dist | 2D | Y | H | M | H | M |
| **TMWC [10]** | U | M | - | N | N | Y | RSSI Conn | Centro | Dist | 2D | Y | L | L | M | H |

A summary of comparison between range-based and range-free schemes

**Net**: Network –**De**: deployment – **Nd**: Node density – **Ob**: obstacle - **An**: Anchor nodes - **Nm**: Nodes mobility - **Ma**: Mobile assisted - **Rae**: Range estimation - **Rac**: Range combination - **Com**: Computational model - **Lc**: localization coordinates - **Sca**: scalability - **Cm**: Communication cost - **Cp**: Computation cost - **Hc**: Hardware cost - **Ac**: Accuracy - **Dist**: Distributed - **Cent**: Centralized - **Multi**: Multilateration - **Trilat** :Trilateration - **Triang**: Triangulation - **Trigon**: Trigonometric - **Conn**: Connectivity **- centro**: centroid - **BPF**: Bayesian Practicle Filter - **LSE:** Least Squares Estimation **- TSE:** Taylor Series Expansion - **MLE:** Maximum Likelihood Estimation - **Y**:Yes - **N**:No - **L**:Low - **M**:Med - **H**:High - **V.H:** Very High - **U**:Uniform – **R**:Random

*A. Accuracy*

Accuracy is the most important key for localization evaluation. Most of the applications need high accuracy.

It is usually defined as the expected Euclidean distance between the location estimate and the actual location of unknown node. Generally, range-based schemes are more accurate than range-free schemes because the range-based schemes use the measurement techniques to calculate the Euclidian distance between a pair of nodes . On the other hand , some range- free localization schemes estimate the Euclidian distance between a pair of sensor nodes, considering the shortest path between them [72]. However, this is only right to such cases where shortest paths are similar to a straight line. and this is normally not valid in real environment.

Moreover, hybrid localization schemes whether range-based or range-free provide better accuracy than the fully localization scheme.

Usually, measurement noise for different measurement kinds arrives from different sources. Therefore errors in the position estimate for each measurement kind are at least partially independent. This independence between different measurement kinds can be exploited by data fusion techniques [73] to create estimators that have better accuracy than estimators based on single measurement kinds.

As we see in table 1, hybrid-range- free scheme like IHLA [30] performs more accurate estimation than fully-range- free scheme like CA [11]. Similarly, the hybrid -range- based scheme EATL [65] performs more accurate estimation than fully-range- based scheme like RTSA [55].

The accuracy of most localization schemes is reduced in obstructed environment [74].this is due to the existence of obstacles, which obstruct the line of sight between sensor nodes. We notice that for the range-based schemes, the accuracy relies largely on the physical sources of localization errors. The physical sources are represented by a wide range of noises and quantization losses of range measurements. For instance, the scheme COLA [32] is sensitive to environmental effects.

The deployment strategy of the sensor nodes (how sensor nodes are scattered) has also a great impact on the localization accuracy. Most localization schemes achieve a good accuracy in uniformly distributed WSN compared to randomly distributed networks. Moreover, some planned deployment of nodes as considered in [75] improves the accuracy and the coverage of the proposed method considerably compared to centroid based method.

In anchor-based schemes, the placement and density of anchor nodes in the network affect the localization accuracy. Existing work finds that when anchor nodes are placed in a convex hull around the network, a higher localization accuracy can be achieved . Installing additional anchor nodes in the center of the network is also helpful.

There are some schemes that perform well only in a network with a high anchor density while someone needs only few anchors. As the case of APIT [21], which works well when there are many anchor nodes and it has low computation and communication cost . On the other hand, mobile assisted localization schemes like ADAL [26], MABT [47] used mobile anchor node instead of static anchor nodes and can achieve better localization accuracy. These schemes are suitable to obstructed environments, because anchor node can move around obstacles, and can be also used to enlarge the sensing coverage area. However, Trajectory design is an overlooked problem with mobile assisted localization in wireless sensor networks.

Many localization schemes are sensitive to node density (amount of localizing nodes per area unit). For example, Hop-count-based algorithms as CDV-Hop [17] scheme usually need high node density so that the hop count approximation for distance is precise. Similarly, algorithms that depend on the range measurement techniques cannot apply to low density because the connectivity failures due to higher distance between sensor nodes. Thus when creating or selecting a localization scheme, it is necessary to consider its requirement on node density. On the other hand, the use of the node mobility like ATPA [62] scheme can help overcome several typical problems in WSNs, such as, low node density, obstacle, network connectivity or fault tolerance.

Usually, centralized schemes provide more accurate position estimates than distributed ones. because the distributed schemes may involve a loss of information due to an incomplete network map and parallel computations.

An example of the centralized scheme is MDS-MAP [27] and another example of the distributed scheme is CDV-Hop [17].

*B. Overhead*

The overhead of localization scheme refers to several items, including, computation and communication cost, number of anchor nodes, processing time, energy consumption, hardware required by each node, etc. This paper tried to introduce the more important criteria to compare the localization schemes, we quote them below:

- ✓ **Communication cost**

The energy efficiency is critical to WSN. A sensor node consumes maximum energy in data communication. This involves both data transmission and reception.
The distributed schemes require collaborations among neighboring nodes. In particular, the scheme based on hop-count as the case of CDV-Hop [17] requires high communication cost. Thus, the localization scheme should minimize the amount of node to node communication.

An adequate communication range can be used to conserve communication cost. On the other hand, generally centralized algorithms incur a high communication costs to move data back to the base station. However, centralized schemes are more accurate than the distributed schemes.

- ✓ **Computation cost**

The processor is the second principal source of exhausting battery life. Energy consumption in data communication is much higher compared to data processing. Centralized localization scheme like MDS-MAP[27] requires range measurements from all the nodes. This is expensive in terms of forwarding the measurements to the processing node and solving the high dimension matrix. On the other hand, in three dimensional scenarios, more efficient data structure and modeling techniques are required for location estimation in a 3D space. thus, computation complexity increases drastically, compared to two dimensions. For example, the scheme NCA [12] requires more computation cost than the scheme CA [11].

The computation complexity of the hybrid schemes grows with the variety and the amount of network data(range measurements, topology informations) used in the location estimation.

Moreover, most hybrid schemes whether range-based or range-free require more computation cost compared with the fully scheme . An example of fully-range-free scheme is CA [11] and another example of hybrid-range-free is IHLA [30].

- ✓ *Hardware cost*

Hardware cost includes the node density, the anchor density and measurement equipments.
In table 1, we consider only the measurement equipment to present the hardware cost. Generally, expensive equipments provide more accurate measurements. A range-based scheme usually requires more hardware than range-free scheme in order to measure the distance between sensor nodes. As the case of the range-free scheme CA [11] and the range-based scheme ALAR [42].
Moreover, some hybrid-range-based schemes are much complex than fully-range-based schemes and involve hardware complexity specially when using two different measurement techniques which require two different hardware. We cite as example, hybrid-range-based scheme EATL [65] and fully-range-based scheme AHLoS [4].

*C. Scalability*

The scalability is an important factor to validate the localization scheme. It ensures suitable estimation of localization when the network or deployment area gets larger.

We notice that for range-based schemes, the location performance reduces when the distance between the sender and receiver increases. On the other hand, in dense network wireless signal channels may become congested, and more complex communication infrastructure may be required. Centralized schemes usually provide more accurate position estimates than distributed ones. Centralized schemes aggregate all measurements and input data at a base station to carry out processing. However, they have a single point of failure, and need a high communication cost. Further, the sensor nodes located close to the base station will drain their energy earlier than others because of funneling effect. Thus, the centralized approaches suffer from scalability. We cite the example of MDS-MAP [27]. In contrast distributed scheme is suitable for large scale networks. This is the case of the scheme IMDS-MAP [29].

## 4. CONCLUSION AND FUTURE WORK

Localization in wireless sensor network is a hot area of research that has been addressed through many proposed schemes. Based on the dependency of the range measurements theses proposal schemes are classified into two major categories: range-based schemes and range-free schemes. However, it is difficult to classify hybrid schemes which combine different methods based on connectivity information and/or range measurement techniques as range-based or range-free schemes. In this paper we make the classification of any localization schemes easy, where range-based schemes and range-free schemes are divided into two types: fully schemes and hybrid schemes. Furthermore, this classification is proposed also to help in comparing localization schemes in terms of accuracy. In particular, between the schemes of the same category either for range- based or for range- free categories.

Although WSNs are a current area of research, there are already various localization schemes, each with an emphasis on specific scenario and/or application. In this paper, we analyze and compare the more representative localization scheme, this comparison was based basically in the following parameters: network assumptions (deployment, Node density, existence of obstacle, existence of anchor node, nodes mobility and mobile assisted), localization process (range estimation, range combination, computational model and localization coordinates), and design goal (scalability, overhead and accuracy).

Among all studied schemes, this comparative analysis conducted us to conclude that each algorithm has its own typical features and none is absolutely the best. On the whole, the range-based methods are either expensive with respect to hardware cost, or susceptibility to environmental noises and dynamics. In contrast, the range-free methods are imprecise and easily affected by node density. On the other hand, hybrid localization scheme provides better accuracy than any single localization scheme. However, it is more complex and need more computation time. Furthermore, the significance of this comparative study relies in offering other authors the possibility of utilizing this analysis to identify the localization schemes which best suits their particular problem.

As we known accuracy is the most important key for localization performance. Among the schemes analyzed in this paper, hybrid schemes look promising. But it still suffers from the time of execution needed for the calculation. Optimization algorithms for accelerating this time is perspective making this scheme an effective solution for the localization in wireless sensor networks. Furthermore, the development of new combination between the range measurement techniques and/ or between range measurement techniques and connectivity methods for different application highly motivated the study in this direction.

International Journal of Computer Networks & Communications (IJCNC) Vol.5, No.6, November 2013

## Authors

**Asma Mesmoudi,**  received her  Master degree in Telecommunication  from the University of Tlemcen (Algeria) in 2011 . Member of STIC laboratory in the same university, her research interests include wireless  sensor networks, wireless  localization techniques.

**Mohammed Feham**, received his PhD in Engineering in optical and microwave communications from the University of Limoges (France) in 1987, and his PhD in science from the university of Tlemcen (Algeria) in1996. Since 1987, he has been Assistant Professor and Professor of Microwave, Communication Engineering and Telecommunication network. His research interests cover telecommunication systems and mobile networks.

**Nabila Labraoui,** is an associate professor in computer engineering at the university of Tlemcen. She received her Ph.D. in computer engineering from the University of Tlemcen. Her current research interests include wireless ad hoc sensor networks, Network Security, Localization and Trust management for distributed and mobile systems.